\documentstyle[12pt,epsf]{article}

\def\ep{\epsilon}

\def\spose#1{\hbox to 0pt{#1\hss}}
\def\lsim{\mathrel{\spose{\lower 3pt\hbox{$\mathchar"218$}}
 \raise 2.0pt\hbox{$\mathchar"13C$}}}
\def\gsim{\mathrel{\spose{\lower 3pt\hbox{$\mathchar"218$}}
 \raise 2.0pt\hbox{$\mathchar"13E$}}}

\begin{document}

\begin{titlepage}

\begin{flushright}
CERN-TH/97-304\\
FTUV/97-48\\
IFIC/97-79\\
hep-ph/9711238
\end{flushright}

\vspace{1.0cm}
\begin{center}
\Large\bf
Two-loop hybrid renormalization of local\\
dimension-4 heavy--light operators
\end{center}

\vspace{0.5cm}
\begin{center}
G. Amor\'os\\[0.1cm]
{\sl Departament de F\'\i sica Te\`orica and IFIC\\
Centre Mixte Universitat de Val\`encia-CSIC\\
E-46100 Burjassot, Val\`encia, Spain}\\
\vspace{0.5cm}
and\\
\vspace{0.5cm}
M. Neubert\\[0.1cm]
{\sl Theory Division, CERN, CH-1211 Geneva 23, Switzerland}
\end{center}

\vspace{1.5cm}
\begin{abstract}
\vspace{0.2cm}\noindent
The renormalization of local dimension-4 operators containing a heavy
and a light quark field at scales below the heavy-quark mass is
discussed, using the formalism of the heavy-quark effective theory. The
anomalous dimensions of these operators and their mixing are calculated
to two-loop order. Some phenomenological applications are briefly
discussed.
\end{abstract}

\vspace{1.0cm}
\centerline{(Submitted to Physics Letters B)}

\vspace{3.0cm}
\noindent
CERN-TH/97-304\\
November 1997

\end{titlepage}

\section{Introduction}

Local operators containing both heavy and light quark fields exhibit an
interesting behaviour under renormalization at scales below the
heavy-quark mass $m_Q$. Large logarithms of the type $\alpha_s
\ln(m_Q/\mu)$ arise from the exchange of gluons that are ``hard'' with
respect to the light quark but ``soft'' with respect to the heavy
quark. Since such gluons see the heavy quark as a static colour source,
the large logarithms can be summed to all orders in perturbation theory
using an effective theory for static heavy quarks, the so-called
heavy-quark effective theory (HQET) \cite{EiHi}--\cite{review}. In the
HQET, the 4-component heavy-quark field $Q(x)$ is replaced by a
velocity-dependent 2-component field $h_v(x)$ satisfying
$\rlap{/}v\,h_v=h_v$, where $v$ is the velocity of the hadron
containing the heavy quark. Because of the particular hierarchy of the
mass scales involved, the renormalization of heavy--light operators in
the HQET is called ``hybrid'' renormalization. Operators in the
effective theory have a different evolution than in usual QCD. For
instance, whereas the vector current $\bar q\,\gamma^\mu Q$ is
conserved in QCD (i.e.\ its anomalous dimension vanishes), the
corresponding current $\bar q\,\gamma^\mu h_v$ in the HQET has a
nontrivial anomalous dimension \cite{KoRa}--\cite{PoWi}, which governs
the evolution for scales below the heavy-quark mass. It has been
calculated at the two-loop order in Refs.~\cite{JiMu}--\cite{Gim}. The
renormalization of higher-dimensional heavy--light operators is known
only at the leading logarithmic order \cite{FaGr}--\cite{Groz}. Here we
shall generalize these results and perform the calculation of the
anomalous dimension matrix of local dimension-4 operators at the
two-loop order. This calculation requires the evaluation of two-loop
tensor integrals in the HQET that are infrared (IR) singular when one
of the external lines is taken on-shell. A general algorithm to
calculate these integrals has been developed recently \cite{ABN}.

The matrix elements of local dimension-4 heavy--light operators appear
at order $1/m_Q$ in the heavy-quark expansion of heavy-meson decay
constants \cite{subl} and weak transition form factors \cite{Burd}. As
such, they play an important role in many phenomenological applications
of the HQET. The theoretical predictions for weak decay form factors
involve operator matrix elements renormalized at the scale $m_Q$. Our
results can then be used to rewrite these matrix elements in terms of
ones renormalized at a scale $\mu\ll m_Q$, which may be identified with
the scale at which a nonperturbative evaluation of these matrix
elements is performed. Details of such applications will be discussed
elsewhere.

\section{Operator renormalization}

We start by constructing, in the HQET, a basis of local dimension-4
operators containing a heavy and a light quark field. Since ultimately
our interest is in the matrix elements of these operators between
physical hadron states, it is sufficient to consider gauge-invariant
operators that do not vanish by the equations of motions \cite{Simm}.
Then, in the limit where the light-quark mass is set to zero, any
dimension-4 operator must contain a covariant derivative acting on one
of the quark fields. A basis of such operators, which closes under
renormalization, is provided by\footnote{We use the notation
$iD=i\partial+g_s A$ and $(iD)^\dagger=-i\overleftarrow{\partial}+g_s
A$.}
\begin{eqnarray}
   O_1 &=& \bar q\,\Gamma\,i D_\alpha h_v \,, \nonumber\\
   O_2 &=& \bar q\,(i D_\alpha)^\dagger\,\Gamma\,h_v \,,
    \nonumber\\
   O_3 &=& \bar q\,(iv\!\cdot\!D)^\dagger v_\alpha\Gamma\,h_v \,,
    \nonumber\\
   O_4 &=& \bar q\,(iv\!\cdot\!D)^\dagger\gamma_\alpha\rlap/v\,
    \Gamma\,h_v \,,
\label{ops}
\end{eqnarray}
where $\Gamma$ represents an arbitrary Dirac matrix, which may or may
not contain the Lorentz index $\alpha$. The Feynman rules of the HQET
ensure that no Dirac matrices appear on the right-hand side of
$\Gamma$. Moreover, for $m_q=0$ only an even number of $\gamma$
matrices can appear on the left-hand side of $\Gamma$.

The equations of motion, $iv\cdot D\,h_v=0$ and $\bar q\,
(i\rlap{\,/}D)^\dagger=0$, imply that between physical states the
operators $(O_1-O_2)$, $O_3$ and $O_4$ can be replaced by total
derivatives of some lower-dimensional operators, which are renormalized
multiplicatively and irrespective of their Dirac structure. The
additional power of the external momentum carried by the operators does
not affect the ultraviolet (UV) behaviour of loop diagrams. Moreover,
the reparametrization invariance of the HQET \cite{LuMa,Chen} ensures
that also the multiplicative renormalization of the operator $O_1$ is
determined by the same $Z$ factor that governs the renormalization of
dimension-3 operators \cite{MN94}. However, there is a nontrivial
mixing between $O_1$ and the other three operators.

We define a matrix ${\bf Z}$ of renormalization constants, which absorb
the UV divergences in the matrix elements of the bare operators, by the
relation $O_i = \sum_j\,{\bf Z}_{ij}\,O_{j,{\rm bare}}$. Then the
matrix \boldmath$\gamma$\unboldmath\ of the anomalous dimensions, which
govern the scale dependence of the renormalized operators, is given by
\begin{equation}
   \mbox{\boldmath$\gamma$\unboldmath}
   = - \frac{{\rm d}{\bf Z}}{{\rm d}\ln\mu}\,{\bf Z}^{-1} \,.
\end{equation}
From the arguments presented above, it follows that the matrix ${\bf
Z}$ has the structure
\begin{equation}
   {\bf Z} = \left( \begin{array}{cccc}
   ~Z_1~ & Z_2 & ~Z_3~ & ~Z_4~ \\
   0 & Z_1+Z_2 & Z_3 & Z_4 \\
   0 & 0 & Z_1 & 0 \\
   0 & 0 & 0 & Z_1
   \end{array} \right) \,,
\label{Zmat}
\end{equation}
where $Z_1$ is the renormalization constant of local dimension-3
currents, which is known at the two-loop order \cite{JiMu}--\cite{Gim}.
Here we will calculate the remaining entries in the matrix ${\bf Z}$
with the same accuracy.

In a minimal subtraction scheme, the renormalization constants are
defined to remove the $1/\ep$ poles arising in the calculation of the
bare Green functions with insertions of the operators $O_i$ in
dimensional regularization, i.e.\ in $d=4-2\ep$ space-time dimensions.
Hence,
\begin{equation}
   {\bf Z} = {\bf 1} + \sum_{k=0}^\infty\,\frac{1}{\ep^k}\,
   {\bf Z}^{(k)} \,.
\end{equation}
The requirement that the anomalous dimension matrix be finite in the
limit $\ep\to 0$ implies the relations \cite{Flor}
\begin{eqnarray}
   \mbox{\boldmath$\gamma$\unboldmath} &=& 2\alpha_s
    \frac{\partial{\bf Z}^{(1)}}{\partial\alpha_s}
    = - 2\alpha_s
    \frac{\partial\big({\bf Z}^{-1}\big)^{(1)}}{\partial\alpha_s}
    \,, \nonumber\\
   \alpha_s \frac{\partial{\bf Z}^{(2)}}{\partial\alpha_s}
   &=& \alpha_s \frac{\partial{\bf Z}^{(1)}}{\partial\alpha_s}
    \left( {\bf Z}^{(1)} + \frac{\beta(\alpha_s)}{\alpha_s} \right) \,,
\label{magic}
\end{eqnarray}
where $\beta(\alpha_s)={\rm d}\alpha_s/{\rm d}\ln\!\mu^2$ is the
$\beta$ function. The first equation shows that the anomalous dimension
matrix can be obtained from the coefficient of the $1/\ep$ pole in
${\bf Z}$, whereas the second one implies a nontrivial constraint on
the coefficient of the $1/\ep^2$ pole, arising at two-loop and higher
order.

Previous authors have calculated the renormalization constants $Z_i$
appearing in (\ref{Zmat}) at the one-loop order, finding that
\cite{FaGr,MN94}
\begin{equation}
   Z_i = \delta_{1i} + \frac{C_F\alpha_s}{4\pi\ep}
   \left( -\frac 32 \,, \frac 32 \,, -1 \,, -\frac 12
   \right) + O(\alpha_s^2) \,; \quad i=1,2,3,4 \,.
\label{Z1loop}
\end{equation}
This information is sufficient to reconstruct the matrix ${\bf Z}$ at
order $\alpha_s$. From the second relation in (\ref{magic}), it then
follows that
\begin{equation}
   Z_i^{(2)} = \left( \frac{\alpha_s}{4\pi} \right)^2
   \left( -\frac 34 C_F^2 - \frac{11}{6} C_F C_A
   + \frac 23 C_F T_F\,n_f \right)\,\left( -\frac 32 \,, \frac 32 \,,
   -1 \,, -\frac 12 \right) + O(\alpha_s^3) \,.
\label{ep2terms}
\end{equation}
Here $C_A=N$, $C_F=\frac 12 (N^2-1)/N$ and $T_F=\frac 12$ are the
colour factors for an $SU(N)$ gauge group, and $n_f$ is the number of
light-quark flavours. This relation will provide a check on our
two-loop results.

The equations of motion impose two additional conditions on the
renormalization constants. Substituting $\Gamma=v^\alpha\Gamma'$ or
$\Gamma=\gamma^\alpha\Gamma'$ in the bare operator $O_1$ in
(\ref{ops}), and using that between physical states $\bar q\,\Gamma'\,i
v\cdot D h_v=0$, whereas $\bar q\,i\rlap{\,/}D\Gamma' h_v
=i\partial_\alpha\,[\bar q\,\gamma^\alpha\Gamma' h_v]$ is renormalized
multiplicatively, we find that
\begin{eqnarray}
   ({\bf Z}^{-1})_{12} + ({\bf Z}^{-1})_{13}
   + ({\bf Z}^{-1})_{14} &=& 0 \,, \nonumber\\
   ({\bf Z}^{-1})_{13} - 2(1-\ep) ({\bf Z}^{-1})_{14}\Big|_{\rm poles}
   &=& 0 \,,
\label{eom}
\end{eqnarray}
where it is understood that in the second relation all pole terms
proportional to $1/\ep^k$ must cancel. In terms of the constants $Z_i$,
we obtain the conditions
\begin{eqnarray}
   &&Z_2^{(k)} + Z_3^{(k)} + Z_4^{(k)} = 0 \,;\quad k\ge 1 \,,~~
    \nonumber\\
   &&Z_3^{(1)} - 2 Z_4^{(1)}\,\Big[ 1 + 2 Z_1^{(1)} + Z_2^{(1)} \Big]
    + 2 Z_4^{(2)} = 0 \,,~~
\label{eomb}
\end{eqnarray}
which will provide a highly nontrivial check on our two-loop results.
Relations (\ref{eom}) allow to calculate two of the three constants
$Z_2$, $Z_3$ and $Z_4$ in terms of the third one. The fact that the
calculation of one of these constants suffices to reconstruct the full
matrix ${\bf Z}$, using that $Z_1$ is fixed by reparametrization
invariance, has been observed in Ref.~\cite{MN94}; however, the
relations derived there are not fully correct beyond the one-loop
order.

\begin{figure}
\epsfxsize=13cm
\centerline{\epsffile{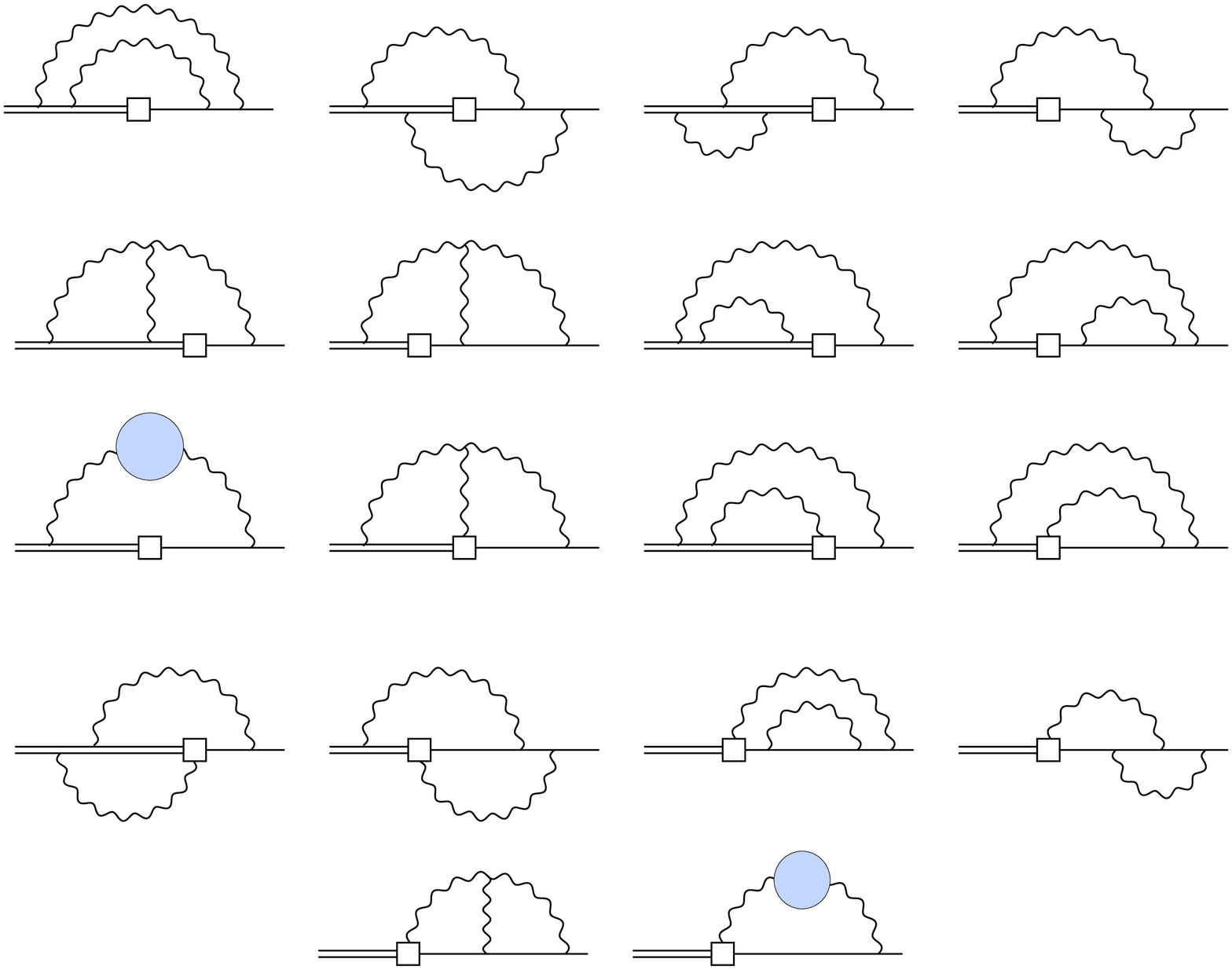}}
\vspace{-0.4cm}
\centerline{\parbox{12cm}{\caption{\label{fig:2loop}
\small\sl Two-loop diagrams contributing to the calculation of the
renormalization constants $Z_i$. The shaded circles represent one-loop
insertions of the gluon self-energy.}}}
\end{figure}

\section{Two-loop calculation}

To obtain the renormalization constants $Z_i$ at order $\alpha_s^2$, we
calculate the insertions of the operator $O_1$ into the amputated Green
function with a heavy and a light quark to two-loop order. The relevant
diagrams are shown in Figure~\ref{fig:2loop}. We need to keep terms
linear in the momentum $p$ of the light quark, as only those contribute
to the renormalization constants $Z_2$, $Z_3$ and $Z_4$. However, for
completeness we will also keep terms linear in the heavy-quark momentum
$k_\alpha$, which contribute to $Z_1$. In that way we will reproduce
the known two-loop result for $Z_1$ as a check of our calculation.
Because the pole parts are polynomial in the external momenta, we can
first take a derivative with respect to $p$ and $k$ and then set $p=0$,
so that all integrals are of propagator type and depend on the single
variable $\omega=v\cdot k$. However, this method fails for some of the
diagrams, for which setting $p=0$ after differentiation leads to IR
divergences. In these cases, we apply a variant of the so-called $R^*$
operation \cite{Che}, which compensates these IR poles by a
construction of counterterms for the IR divergent subgraphs.

\begin{figure}
\epsfxsize=10cm
\centerline{\epsffile{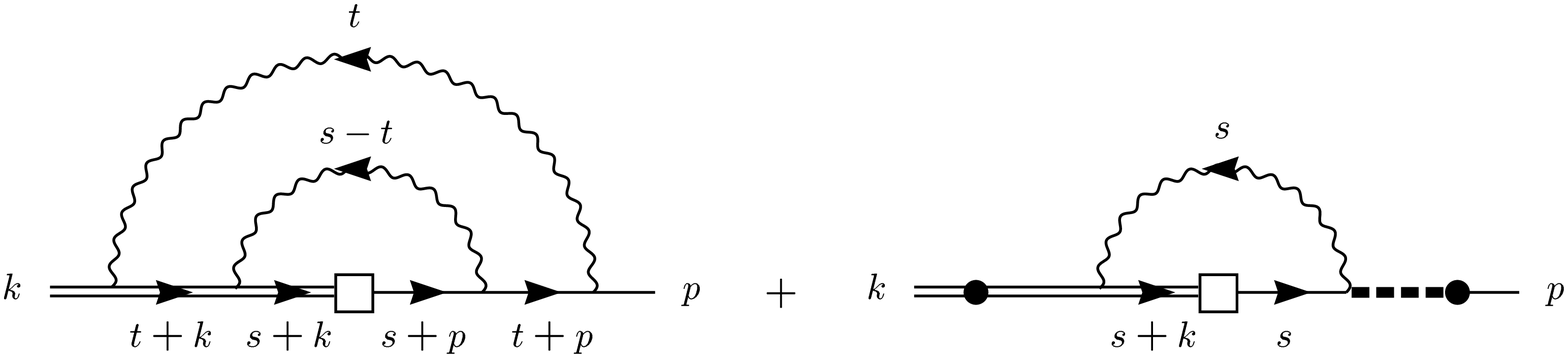}}
\vspace{-1cm}
\centerline{\parbox{12cm}{\caption{\label{fig:rstar}
\small\sl Schematic representation of the $R^*$ operation. The black
dots represent the original vertices, the dashed line the IR
counterterm.}}}
\end{figure}

Consider, as an example, the first diagram in Figure~\ref{fig:2loop}.
Its contribution is proportional to the integral
\begin{equation}
   D_1 = \int\!{\rm d}^ds\,{\rm d}^dt\,
   \frac{(s+k)_\alpha\,(t+p)_\beta\,(s+p)_\gamma}
        {t^2 (t+p)^2 (s+p)^2 (s-t)^2
         (v\!\cdot\!s+\omega)(v\!\cdot\!t+\omega)} \,.
\end{equation}
When linearizing this expression in $p$, we encounter an IR divergence
from the region $t\to 0$, which can be removed by adding and
subtracting the IR subtraction term
\begin{equation}
   D_1^{\rm IR} = \int\!{\rm d}^ds\,{\rm d}^dt\,
   \frac{(t+p)_\beta\,s_\alpha\,s_\gamma}
        {t^2 (t+p)^2 \big(s^2\big)^2(v\!\cdot\!s+\omega)\,\omega} \,.
\end{equation}
The difference $D_1-D_1^{\rm IR}$ can be evaluated using a naive
linearization in $p$, since the behaviour of the IR subtraction term
for $t\to 0$ is the same as that of the original integral. However,
because the expansion of $D_1^{\rm IR}$ involves tadpole integrals that
vanish in dimensional regularization, the difference $(D_1-D_1^{\rm
IR})_{\rm linearized}$ coincides with the naive linearization of the
original expression for $D_1$. The contribution $D_1^{\rm IR}$, which
is necessary to subtract the IR subdivergence of the original diagram,
factorizes into an IR counter\-term (the $t$ integral) and the original
diagram with the lines of the IR sensitive subgraph removed, and with
$t$ and $p$ set to zero. The construction just described is
schematically shown in Figure~\ref{fig:rstar}, where the dashed line
represents the IR counterterm
\begin{equation}
   \int\frac{{\rm d}^dt}{(2\pi)^d}\,
   \frac{\gamma^\beta(t+p)_\beta}{t^2 (t+p)^2}
   = \frac{i\rlap{\,/}p}{(4\pi)^{d/2}}\,(-p^2)^{-\ep}\,
   \frac{\Gamma(d/2)\,\Gamma(d/2-1)\,\Gamma(2-d/2)}{\Gamma(d-1)} \,.
\end{equation}

The remaining two-loop tensor integrals are of the general form
\begin{eqnarray}
   &&\int\mbox{d}^ds\,\mbox{d}^dt\,
   \left( \frac{\omega}{v\!\cdot\!s+\omega} \right)^{\alpha_1}
   \left( \frac{\omega}{v\!\cdot\!t+\omega} \right)^{\alpha_2}
   \frac{s_{\mu_1}\dots s_{\mu_n} t^{\nu_1}\dots t^{\nu_m}}
    {\big(-s^2\big)^{\alpha_3} \big(-t^2\big)^{\alpha_4}
     \big[-(s-t)^2\big]^{\alpha_5}}
    \nonumber\\
   &&\quad \equiv
    -\pi^d (-2\omega)^{2(d-\alpha_3-\alpha_4-\alpha_5)+n+m}\,
    I_{\mu_1\dots\mu_n}^{\nu_1\dots\nu_m}(v;\{\alpha_i\}) \,,
    \nonumber\\[0.3cm]
   &&\int\mbox{d}^ds\,\mbox{d}^dt\,
   \frac{s_{\mu_1}\dots s_{\mu_n} t^{\nu_1}\dots t^{\nu_m}}
    {\big(-s^2\big)^{\alpha_1} \big(-t^2\big)^{\alpha_2}
     \big[-(s-p)^2\big]^{\alpha_3} \big[-(t-p)^2\big]^{\alpha_4}
     \big[-(s-t)^2\big]^{\alpha_5}}
    \nonumber\\
   &&\quad \equiv -\pi^d (-p^2)^{(d-\alpha_1-\alpha_2-\alpha_3
    -\alpha_4-\alpha_5)}\,
    R_{\mu_1\dots\mu_n}^{\nu_1\dots\nu_m}(p;\{\alpha_i\}) \,.
\end{eqnarray}
Using the method of integration by parts \cite{BrGr,Chet}, we obtain
the recurrence relations
\begin{eqnarray}
   &&\left[ (d-\alpha_1-\alpha_3-2\alpha_5+n)
    + \alpha_3\,{\bf 3^+} ({\bf 4^-} - {\bf 5^-})
    + \alpha_1\,{\bf 1^+ 2^-} \right]\,
    I_{\mu_1\dots\mu_n}^{\nu_1\dots\nu_m}(v;\{\alpha_i\}) \nonumber\\
   &&\quad = \sum_{j=1}^n\,
    I_{\mu_1\dots[\mu_j]\dots\mu_n}^{\mu_j\nu_1\dots\nu_m}
    (v;\{\alpha_i\}) \,, \nonumber\\[0.3cm]
   &&\left[ (d-\alpha_1-\alpha_3-2\alpha_5+n)
    + \alpha_3\,{\bf 3^+} ({\bf 4^-} - {\bf 5^-})
    + \alpha_1\,{\bf 1^+} ({\bf 2^-} - {\bf 5^-}) \right]\,
    R_{\mu_1\dots\mu_n}^{\nu_1\dots\nu_m}(p;\{\alpha_i\}) \nonumber\\
   &&\quad = \sum_{j=1}^n\,
    R_{\mu_1\dots[\mu_j]\dots\mu_n}^{\mu_j\nu_1\dots\nu_m}
    (p;\{\alpha_i\}) \,,
\end{eqnarray}
which allow us to express any two-loop integral in terms of degenerate
integrals, which have $\alpha_2=0$, $\alpha_4=0$ or $\alpha_5=0$. Here
${\bf 1^+}$ is an operator raising the index $\alpha_1$ by one unit
etc., and $[\mu_j]$ means that this index is omitted. The degenerate
integrals can be related in a straightforward way to products of
one-loop tensor integrals \cite{ABN}. Using this technique, we have
calculated the pole parts of the two-loop diagrams in the
`t~Hooft--Feynman gauge. The results are summarized in the first three
columns of Table~\ref{tab:2loop}. The renormalization scale $\mu$ is
introduced by the replacement of the bare coupling constant with the
renormalized one through the relation $g_s^{\rm bare}=\bar\mu^\epsilon
Z_g\,g_s$, with $\bar\mu=\mu\,e^{\gamma_E/2} (4\pi)^{-1/2}$ in the
$\overline{\mbox{MS}}$ scheme.

\begin{table}[t]
\caption{\label{tab:2loop}
\small\sl Sum of two-loop and counterterm contributions in units of
$(\alpha_s/4\pi)^2$}
\vspace{0.4cm}
\begin{tabular}{|l|l|c|c|c|c|c|}\hline\hline
\multicolumn{2}{|l|}{Structure}
 & \rule[-0.35cm]{0cm}{0.95cm}
  $\left( \frac{-2\omega}{\mu} \right)^{-4\ep}$
  \rule[-0.35cm]{0cm}{0.95cm}
 & $\!\left( \frac{-2\omega}{\mu} \right)^{-2\ep}\!
    \left( \frac{-p^2}{\mu^2} \right)^{-\ep}\!\!$
 & $\left( \frac{-p^2}{\mu^2} \right)^{-2\ep}$
 & $\left( \frac{-2\omega}{\mu} \right)^{-2\ep}$
 & $\left( \frac{-p^2}{\mu^2} \right)^{-\ep}$ \\
\hline
$p_\alpha$ \rule{0cm}{0.5cm} & $C_F^2$ &
 $-\frac{3}{8\ep^2}+\left(1-\frac{\pi^2}{3}\right)\frac{1}{\ep}$ &
 $-\frac{1}{2\ep^2}-\frac{1}{\ep}$ &
 $\frac{1}{2\ep^2}+\frac{9}{4\ep}$ & $\frac{5}{4\ep^2}$ &
 $-\frac{1}{2\ep^2}-\frac{1}{\ep}$ \\[0.15cm]
 & $C_F C_A$ & $\!-\frac{2}{3\ep^2}
  +\left(-\frac{49}{36}+\frac{\pi^2}{12}\right)\frac{1}{\ep}\!$ &
 $-\frac{1}{2\ep^2}-\frac{1}{\ep}$ &
 $\!-\frac{19}{12\ep^2}-\frac{151}{18\ep}\!$ & $\frac{11}{6\ep^2}$ &
 $\frac{11}{3\ep^2}+\frac{22}{3\ep}$ \\[0.15cm]
 & $C_F T_F\,n_f\!$ & $\frac{1}{3\ep^2}+\frac{1}{18\ep}$ & 0 &
 $\frac{2}{3\ep^2}+\frac{31}{9\ep}$ & $-\frac{2}{3\ep^2}$ &
 $\!-\frac{4}{3\ep^2}-\frac{8}{3\ep}\!$ \\[0.15cm]
\hline
$v\!\cdot\!p\,v_\alpha$ \rule{0cm}{0.5cm} & $C_F^2$ &
 $\frac{5}{4\ep^2}+\left(\frac{17}{12}+\frac{2\pi^2}{9}\right)
  \frac{1}{\ep}$ & $-\frac{1}{\ep^2}-\frac{5}{\ep}$ & 0 &
 $\!-\frac{3}{2\ep^2}+\frac{1}{2\ep}\!$ &
 $\frac{1}{\ep^2}+\frac{2}{\ep}$ \\[0.15cm]
 & $C_F C_A$ & $\frac{11}{6\ep^2}+\left(-\frac 79
  -\frac{\pi^2}{18}\right)\frac{1}{\ep}$ & 0 & 0 &
 $\!-\frac{11}{3\ep^2}+\frac{11}{3\ep}\!$ & 0 \\[0.15cm]
 & $C_F T_F\,n_f\!$ & $-\frac{2}{3\ep^2}+\frac{5}{9\ep}$ & 0 &
 0 & $\frac{4}{3\ep^2}-\frac{4}{3\ep}$ & 0 \\[0.15cm]
\hline
$v\!\cdot\!p\,\gamma_\alpha\rlap/v$ \hspace{-0.3cm} \rule{0cm}{0.5cm}
 & $C_F^2$ & $\frac{5}{8\ep^2}+\left(\frac 43+\frac{\pi^2}{9}\right)
  \frac{1}{\ep}$ & $-\frac{1}{2\ep^2}-\frac{3}{\ep}$ & 0 &
 $\!-\frac{3}{4\ep^2}+\frac{1}{2\ep}\!$ &
 $\frac{1}{2\ep^2}+\frac{1}{\ep}$ \\[0.15cm]
 & $C_F C_A$ & $\frac{11}{12\ep^2}+\left(\frac{19}{36}
  -\frac{\pi^2}{36}\right)\frac{1}{\ep}$ & 0 & 0 &
 $-\frac{11}{6\ep^2}$ & 0 \\[0.15cm]
 & $C_F T_F\,n_f\!$ & $-\frac{1}{3\ep^2}-\frac{1}{18\ep}$ & 0 &
 0 & $\frac{2}{3\ep^2}$ & 0 \\[0.15cm]
\hline
$k_\alpha$ \rule{0cm}{0.5cm} & $C_F^2$ &
 $\frac{1}{2\ep^2}+\left(-1+\frac{2\pi^2}{3}\right)\frac{1}{\ep}$ &
 0 & 0 & $-\frac{1}{\ep^2}$ & 0 \\[0.15cm]
 & $C_F C_A$ & $\frac{1}{\ep^2}+\left(1-\frac{\pi^2}{6}\right)
  \frac{1}{\ep}$ & 0 & 0 & $-\frac{2}{\ep^2}$ & 0 \\[0.15cm]
 & $C_F T_F\,n_f\!$ & 0 & 0 & 0 & 0 & 0 \\[0.15cm]
\hline\hline
\end{tabular}
\end{table}

The two-loop diagrams in Figure~\ref{fig:2loop} contain subdivergences,
which must be subtracted by UV counterterms. In addition to the
one-loop counterterms for the quark and gluon propagators and vertices,
local operator counterterms are required. To find these, we calculate
at the one-loop order all insertions of the operator $O_1$ into the
amputated Green functions with a non-negative degree of divergence. In
our case, those are the two- and three-point functions with field
content $\bar q h$ and $\bar q h A$. We find that in the
't~Hooft--Feynman gauge the UV divergences of these functions are
removed by the counterterms
\begin{equation}
   {\cal L}_{\rm c.t.} = - C_F\,\frac{\alpha_s}{4\pi\ep}
   \left( O_1 - \frac 32 O_2 + O_3 + \frac 12 O_4 \right)
   - C_A\,\frac{\alpha_s}{4\pi\ep}\,\bar\mu^\epsilon g_s\,
   \bar q\,\Gamma A_\alpha h_v \,.
\end{equation}
Since the two-loop calculation was performed off-shell, a
gauge-dependent operator has to be included in addition to the
operators $O_i$ introduced in (\ref{ops}). However, there is no need to
include operators that vanish by the equations of motion \cite{Simm}.
The results for the sum of all counterterm contributions are summarized
in the last two columns of Table~\ref{tab:2loop}. When these
contributions are added to the result for the sum of the two-loop
diagrams, all nonlocal $1/\epsilon$ divergences proportional to
$\ln(-2\omega/\mu)$ and $\ln(-p^2/\mu^2)$ cancel. This is a nontrivial
check of our calculation.

The sum of the two-loop diagrams plus their counterterms determines the
two-loop coefficients in the products $-Z_h^{1/2} Z_q^{1/2} Z_i$. To
obtain the results for the renormalization constants at order
$\alpha_s^2$, we have to account for wave-function renormalization of
the external quark fields. Using the one-loop expressions for the $Z_i$
factors in (\ref{Z1loop}), as well as the known one- and two-loop (for
$Z_1$) wave-function renormalization constants of the quark fields
\cite{Egor,BGS}, we obtain the final results:
\begin{eqnarray}
   Z_1 &=& 1 - \frac{3 C_F\alpha_s}{8\pi\ep} \Bigg\{
    1 + \frac{\alpha_s}{4\pi} \Bigg[
    C_F \left( -\frac{3}{4\ep} - \frac{5}{12} + \frac{4\pi^2}{9}
    \right) \nonumber\\
   &&\qquad\mbox{}+ C_A \left( -\frac{11}{6\ep}
    + \frac{49}{36} - \frac{\pi^2}{9}
    \right) + T_F\,n_f \left( \frac{2}{3\ep} - \frac{5}{9} \right)
    \Bigg] \Bigg\} \,, \nonumber\\
   Z_2 &=& \frac{3 C_F\alpha_s}{8\pi\ep} \Bigg\{
    1 + \frac{\alpha_s}{4\pi} \Bigg[
    C_F \left( -\frac{3}{4\ep} - \frac{5}{6} + \frac{2\pi^2}{9}
    \right) \nonumber\\
   &&\qquad\mbox{}+ C_A \left( -\frac{11}{6\ep}
    + \frac{41}{18} - \frac{\pi^2}{18}
    \right) + T_F\,n_f \left( \frac{2}{3\ep} - \frac{5}{9} \right)
    \Bigg] \Bigg\} \,, \nonumber\\
   Z_3 &=& -\frac{C_F\alpha_s}{4\pi\ep} \Bigg\{
    1 + \frac{\alpha_s}{4\pi} \Bigg[
    C_F \left( -\frac{3}{4\ep} - \frac{13}{12} + \frac{2\pi^2}{9}
    \right) \nonumber\\
   &&\qquad\mbox{}+ C_A \left( -\frac{11}{6\ep}
    + \frac{26}{9} - \frac{\pi^2}{18}
    \right) + T_F\,n_f \left( \frac{2}{3\ep} - \frac{7}{9} \right)
    \Bigg] \Bigg\} \,, \nonumber\\
   Z_4 &=& -\frac{C_F\alpha_s}{8\pi\ep} \Bigg\{
    1 + \frac{\alpha_s}{4\pi} \Bigg[
    C_F \left( -\frac{3}{4\ep} - \frac{1}{3} + \frac{2\pi^2}{9}
    \right) \nonumber\\
   &&\qquad\mbox{}+ C_A \left( -\frac{11}{6\ep}
    + \frac{19}{18} - \frac{\pi^2}{18}
    \right) + T_F\,n_f \left( \frac{2}{3\ep} - \frac{1}{9} \right)
    \Bigg] \Bigg\} \,.
\end{eqnarray}
Our result for $Z_1$ agrees with that derived in
Refs.~\cite{JiMu}--\cite{Gim}. The two-loop results for the remaining
constants $Z_i$ are new. Note that these expressions do indeed obey the
relations in (\ref{ep2terms}) and (\ref{eomb}). This a highly
nontrivial check, which gives us confidence in the correctness of our
results.

The anomalous dimension matrix \boldmath$\gamma$\unboldmath\ governing
the scale dependence and mixing of the renormalized operators can be
obtained using the first relation in (\ref{magic}). Since the matrix
coefficients in the perturbative expansion
\begin{equation}
   \mbox{\boldmath$\gamma$\unboldmath}
   = \mbox{\boldmath$\gamma_0$\unboldmath}\,\frac{\alpha_s}{4\pi}
   + \mbox{\boldmath$\gamma_1$\unboldmath} \left(
   \frac{\alpha_s}{4\pi} \right)^2 + \dots
\end{equation}
have the same texture as the matrix ${\bf Z}$ in (\ref{Zmat}), it is
sufficient to quote the entries of the first row, which we denote as
$\gamma_i$ with $i=1,2,3,4$. We obtain
\begin{eqnarray}
   (\mbox{\boldmath$\gamma_0$\unboldmath})_{1i}\equiv\gamma_{i,0}
   &=& C_F \left( -3 \,, 3 \,, -2 \,, - 1 \right) \,, \nonumber\\
   (\mbox{\boldmath$\gamma_1$\unboldmath})_{1i}\equiv\gamma_{i,1}
   &=& C_F^2 \left( \frac 52 - \frac{8\pi^2}{3} \,,
    -5 + \frac{4\pi^2}{3} \,, \frac{13}{3} - \frac{8\pi^2}{9} \,,
    \frac 23 - \frac{4\pi^2}{9} \right) \nonumber\\
   &&\mbox{}+ C_F C_A \left( -\frac{49}{6}
    + \frac{2\pi^2}{3} \,, \frac{41}{3} - \frac{\pi^2}{3} \,,
    -\frac{104}{9} + \frac{2\pi^2}{9} \,, -\frac{19}{9}
    + \frac{\pi^2}{9} \right) \nonumber\\
   &&\mbox{}+ C_F T_F\,n_f \left( \frac{10}{3} \,,
    -\frac{10}{3} \,, \frac{28}{9} \,, \frac 29 \right) \,.
\label{anomdim}
\end{eqnarray}
We note that $\gamma_2+\gamma_3+\gamma_4=0$ as a consequence of the
first relation in (\ref{eom}), and that the one-loop matrix coefficient
\boldmath$\gamma_0$\unboldmath\ satisfies the simple relation
$\mbox{\boldmath$\gamma_0$\unboldmath}^2= -3 C_F
\mbox{\boldmath$\gamma_0$\unboldmath}$.

\section{Nontrivial basis transformations}

The results derived in this work are sufficient to calculate, at the
two-loop order, the hybrid renormalization of any local dimension-4
operator containing a heavy and a light quark field. However, it some
cases the choice of the operator basis in (\ref{ops}) may not be the
most convenient one. Whereas the transformation between one operator
basis and another is trivial at the one-loop order, it may become
subtle at the next-to-leading order, because in dimensional
regularization the relations between the operators of different bases
may depend on $\epsilon$.

Consider the general case where the operator basis $\{O_i\}$ is
replaced by a new basis $\{Q_j\}$ with a linear relation of the form
$O_i = \sum_j\,{\bf R}_{ij}(\ep)\,Q_j$. Depending on the choice of the
Dirac matrix $\Gamma$, some of the operators $O_i$ may not be
independent, and thus it may happen that the new basis contains less
than four operators. Therefore, in general the transformation matrix
${\bf R}(\ep)$ is a $4\times n$ matrix with $n\le 4$. We introduce an
$n\times 4$ left-inverse of this matrix such that ${\bf L}(\ep)\,{\bf
R}(\ep)={\bf 1}$. It then follows that the $n\times n$ matrix
\boldmath${\bf\cal Z}$\unboldmath\ of renormalization constants in the
new operator basis satisfies the relation
\begin{equation}
   \mbox{\boldmath${\cal Z}$\unboldmath}^{-1} = {\bf 1}
   + {\bf L}(\ep)\left( {\bf Z}^{-1}-{\bf 1} \right)
   {\bf R}(\ep) \Big|_{\rm poles} \,,
\label{Zinv}
\end{equation}
where only pole terms are kept on the right-hand side. The anomalous
dimension matrix in the new basis is determined by the terms of order
$1/\ep$ in \boldmath${\bf\cal Z}$\unboldmath. Expanding the
transformation matrices as ${\bf R}(\ep)=\sum_n {\bf R_n}\ep^n$ and
${\bf L}(\ep)=\sum_n {\bf L_n}\ep^n$, we obtain
\begin{equation}
   \mbox{\boldmath${\cal Z}$\unboldmath}^{(1)}
   = {\bf L_0}\,{\bf Z}^{(1)}\,{\bf R_0}
   + {\bf L_1}\Big[ {\bf Z}^{(2)} - ({\bf Z}^{(1)})^2 \Big]\,{\bf R_0}
   + {\bf L_0}\Big[ {\bf Z}^{(2)} - ({\bf Z}^{(1)})^2 \Big]\,{\bf R_1}
   + \dots \,,
\end{equation}
where the ellipses represent terms that do not contribute at the
two-loop order. Denoting by \boldmath$\bar\gamma$\unboldmath\ the
anomalous dimension matrix in the new operator basis, we obtain, using
(\ref{magic}),
\begin{eqnarray}
   \mbox{\boldmath$\bar\gamma_0$\unboldmath}
   &=& {\bf L_0}\,\mbox{\boldmath$\gamma_0$\unboldmath}\,{\bf R_0} \,,
    \nonumber\\
   \mbox{\boldmath$\bar\gamma_1$\unboldmath}
   &=& {\bf L_0}\,\mbox{\boldmath$\gamma_1$\unboldmath}\,{\bf R_0}
    - {\bf L_1}\Big( \textstyle\frac 12
    \mbox{\boldmath$\gamma_0$\unboldmath}^2
    + \beta_0\mbox{\boldmath$\gamma_0$\unboldmath} \Big)\,{\bf R_0}
    - {\bf L_0}\Big( \frac 12\mbox{\boldmath$\gamma_0$\unboldmath}^2
    + \beta_0\mbox{\boldmath$\gamma_0$\unboldmath} \Big)\,{\bf R_1} \,.
\end{eqnarray}
Alternatively, we may evaluate (\ref{Zinv}) directly to get the inverse
matrix $\mbox{\boldmath${\cal Z}$\unboldmath}^{-1}$, and then use the
first relation in (\ref{magic}) to calculate the anomalous dimension
matrix.

As an example, we discuss the mixing of the local dimension-4 operators
appearing at order $1/m_Q$ in the heavy-quark expansion of the vector
current\footnote{The discussion for the axial vector current is
identical, provided the basis operators are chosen in a convenient way
\protect\cite{MN94}.}
$V^\mu=\bar q\,\gamma^\mu Q$. There are six such operators, which can
be chosen in the form \cite{FaGr,MN94}
\begin{eqnarray}
   Q_1 &=& \bar q\,\gamma^\mu i\rlap{\,/}D h_v \,,\hspace{1.5cm}
   Q_4 = \bar q\,(iv\!\cdot\!D)^\dagger \gamma^\mu h_v \,,
    \nonumber\\
   Q_2 &=& \bar q\,v^\mu i\rlap{\,/}D h_v \,, \hspace{1.52cm}
   Q_5 = \bar q\,(iv\!\cdot\!D)^\dagger v^\mu h_v \,, \nonumber\\
   Q_3 &=& \bar q\,iD^\mu h_v \,, \hspace{1.75cm}
   Q_6 = \bar q\,(i D^\mu)^\dagger\,h_v \,.
\label{vbasis}
\end{eqnarray}
Each of the first three operators has the structure of $O_1$, with the
substitutions $\Gamma=\gamma^\mu\gamma^\alpha$, $v^\mu\gamma^\alpha$
and $g^{\mu\alpha}$, respectively. In all three cases, the
corresponding operators $O_2$, $O_3$ and $O_4$ can be written as linear
combinations of $Q_4$, $Q_5$ and $Q_6$. For instance, the basis
transformation for the case $\Gamma=\gamma^\mu\gamma^\alpha$ is
\begin{equation}
   \left( \begin{array}{c} O_1 \\ O_2 \\ O_3 \\ O_4 \end{array} \right)
   = \left( \begin{array}{cccc}
   ~1~ & 0 & ~0~ & ~0~ \\
   0 & 0 & 0 & 2 \\
   0 & 1 & 0 & 0 \\
   0 & ~2\ep~ & 4(1-\ep) & 0
   \end{array} \right)
   \left( \begin{array}{c} Q_1 \\ Q_4 \\ Q_5 \\ Q_6 \end{array}
   \right) \,.
\end{equation}
Using (\ref{Zinv}) and the constraints (\ref{eom}) imposed by the
equations of motion, we find for the corresponding matrix of
renormalization constants
\begin{equation}
   \mbox{\boldmath${\cal Z}$\unboldmath}_{(1,4,5,6)}
   = \left( \begin{array}{cccc}
   ~Z_1~ & ~2 Z_4~ & ~2 Z_3~ & 2 Z_2 \\
   0 & Z_1 & 0 & 0 \\
   0 & 0 & Z_1 & 0 \\
   0 & Z_4 & Z_3 & Z_1+Z_2
   \end{array} \right) \,,
\end{equation}
where the subscript indicates the operators $Q_j$ whose renormalization
is accomplished by this matrix. Similarly, for the other two cases
$\Gamma=v^\mu\gamma^\alpha$ and $g^{\mu\alpha}$ we find, respectively,
\begin{equation}
   \mbox{\boldmath${\cal Z}$\unboldmath}_{(2,5)}
   = \left( \begin{array}{cc}
   Z_1 & ~0~ \\
   ~0~ & Z_1
   \end{array} \right) \,, \qquad
   \mbox{\boldmath${\cal Z}$\unboldmath}_{(3,4,5,6)}
   = \left( \begin{array}{cccc}
   ~Z_1~ & ~Z_4~ & ~Z_3~ & Z_2 \\
   0 & Z_1 & 0 & 0 \\
   0 & 0 & Z_1 & 0 \\
   0 & Z_4 & Z_3 & Z_1+Z_2
   \end{array} \right) \,.
\end{equation}
Note that the $3\times 3$ submatrix renormalizing the operators $Q_4$,
$Q_5$ and $Q_6$ is the same in $\mbox{\boldmath${\cal
Z}$\unboldmath}_{(1,4,5,6)}$ and $\mbox{\boldmath${\cal
Z}$\unboldmath}_{(3,4,5,6)}$. This provides a nontrivial check of the
formalism. The above results can thus be combined into a $6\times 6$
matrix of renormalization constants for the six vector-current
operators in (\ref{vbasis}). The corresponding anomalous dimension
matrix \boldmath$\gamma_V$\unboldmath\ has the block form \cite{MN94}
\begin{equation}
   \mbox{\boldmath${\gamma_V}$\unboldmath}
   = \gamma_1\,{\bf 1} + \left( \begin{array}{cc}
   {\bf 0} & ~{\bf A} \\
   {\bf 0} & ~{\bf B}
   \end{array} \right) \,,
\end{equation}
where
\begin{equation}
   {\bf A} = \left( \begin{array}{ccc}
   2\gamma_4 & 2\gamma_3 & 2\gamma_2 \\
   0 & 0 & 0 \\
   ~\gamma_4~ & ~\gamma_3~ & ~\gamma_2~
   \end{array} \right) \,,\qquad
   {\bf B} = \left( \begin{array}{ccc}
   0 & 0 & 0 \\
   0 & 0 & 0 \\
   ~\gamma_4~ & ~\gamma_3~ & ~\gamma_2~
   \end{array} \right) \,.
\end{equation}
The exact two-loop expressions for these matrices, which are obtained
from (\ref{anomdim}), are an important result of our work.

\section{Conclusion}

We have calculated, at the two-loop order in the HQET, the
renormalization and mixing of local dimension-4 operators containing a
heavy and a light quark field. Our formalism allows for an arbitrary
Dirac and Lorentz structure of these operators. The relevant two-loop
diagrams have been evaluated with the help of tensor recurrence
relations, once the IR divergent subgraphs have been regulated using IR
counterterms. We find that a set of (up to) four operators closes under
renormalization. The equations of motion imply nontrivial relations
between the anomalous dimensions of these operators, which we have
derived. We have also shown how general basis transformations, which
may depend on the dimensional regulator $\epsilon$, can be implemented
in our approach.

The matrix elements of local dimension-4 heavy--light operators play an
important role in many phenomenological applications of the heavy-quark
expansion. In particular, they appear at order $1/m_Q$ in the
heavy-quark expansion of heavy-meson decay constants, and of the
semileptonic form factors describing, e.g., $\bar B\to\pi\,\ell\,
\bar\nu$ and $\bar B\to\rho\,\ell\,\bar\nu$ transitions. Our results
are an important step towards a full next-to-leading order analysis of
these quantities. However, still missing is the two-loop mixing of some
nonlocal operators with the local operators considered here. This will
be discussed elsewhere \cite{future}.

\vspace{0.3cm}
{\it Acknowledgements:\/}
M.N.\ would like to thank Martin Beneke and Gerhard Buchalla for
helpful discussions. G.A.\ acknowledges a grant from the Generalitat
Valenciana, and partial support by DGICYT under grant PB95-1096. He
also acknowledges the hospitality of the CERN Theory Division, where
part of this research was performed.


\begin{thebibliography}{99}
\parskip=0pt

\bibitem {EiHi}
E. Eichten and B. Hill, Phys.\ Lett.\ B {\bf 234}, 511 (1990); {\bf
243}, 427 (1990).

\bibitem {Geor}
H. Georgi, Phys.\ Lett.\ B {\bf 240}, 447 (1990).

\bibitem {review}
For a review, see: M. Neubert, Phys.\ Rep.\ {\bf 245}, 259 (1994).

\bibitem {KoRa}
G.P. Korchemsky and A.V. Radyushkin, Yad.\ Fiz.\ {\bf 45}, 198 (1987)
[Sov.\ J.\ Nucl.\ Phys.\ {\bf 45}, 127 (1987)].

\bibitem {hybr}
M.A. Shifman and M.B. Voloshin, Yad.\ Fiz.\ {\bf 45}, 463 (1987)
[Sov.\ J.\ Nucl.\ Phys.\ {\bf 45}, 292 (1987)].

\bibitem {PoWi}
H.D. Politzer and M.B. Wise, Phys.\ Lett.\ B {\bf 206}, 681 (1988);
{\bf 208}, 504 (1988).

\bibitem {JiMu}
X. Ji and M.J. Musolf, Phys.\ Lett.\ B {\bf 257}, 409 (1991).

\bibitem {BrGr}
D.J. Broadhurst and A.G. Grozin, Phys.\ Lett.\ B {\bf 267}, 105
(1991).

\bibitem {Gim}
V. Gim\'enez, Nucl.\ Phys.\ B {\bf 375}, 582 (1992).

\bibitem {FaGr}
A.F. Falk and B. Grinstein, Phys.\ Lett.\ B {\bf 247}, 406 (1990).

\bibitem {MN94}
M. Neubert, Phys.\ Rev.\ D {\bf 49}, 1542 (1994).

\bibitem {Groz}
A.G. Grozin and M. Neubert, Nucl.\ Phys.\ B {\bf 495}, 81 (1997).

\bibitem {ABN}
G. Amor\'os, M. Beneke and M. Neubert, Phys.\ Lett.\ B {\bf 401}, 81
(1997).

\bibitem {subl}
M. Neubert, Phys.\ Rev.\ D {\bf 46}, 1076 (1992).

\bibitem {Burd}
G. Burdman, Z. Ligeti, M. Neubert and Y. Nir, Phys.\ Rev.\ D {\bf 49},
2331 (1994).

\bibitem {Simm}
H. Simma, Z.\ Phys.\ C {\bf 61}, 67 (1994).

\bibitem {LuMa}
M. Luke and A.V. Manohar, Phys.\ Lett.\ B {\bf 286}, 348 (1992).

\bibitem {Chen}
Y.-Q. Chen, Phys.\ Lett.\ B {\bf 317}, 421 (1993).

\bibitem {Flor}
E.G. Floratos, D.A. Ross and C.T. Sachrajda, Nucl.\ Phys.\ B {\bf 129},
66 (1977).

\bibitem {Che}
K.G. Chetyrkin and F.V. Tkachov, Phys.\ Lett.\ B {\bf 114}, 340
(1982);\\
K.G.~Chetyrkin and V.A.~Smirnov, Phys.\ Lett.\ B {\bf 144}, 419
(1984).

\bibitem {Chet}
K.G. Chetyrkin and F.V. Tkachov, Nucl.\ Phys.\ B {\bf 192}, 159
(1981).

\bibitem {Egor}
E.S. Egorian and O.V. Tarasov, Theor.\ Math.\ Phys.\ {\bf 41}, 26
(1979).

\bibitem {BGS}
D.J. Broadhurst, N. Gray and K. Schilcher, Z.\ Phys.\ C {\bf 52}, 111
(1991).

\bibitem {future}
G. Amor\'os and M. Neubert, in preparation.

\end{thebibliography}
\end{document}